\documentstyle[12pt,epsf]{article}
\begin{document}
\newcommand{\ad}[1]{\mbox{${#1}^{\dagger}$}}
\newcommand{\kone}[1]{\mbox{$|{#1}\rangle$}}
\newcommand{\ktwo}[2]{\mbox{$|{#1}\, ,{#2}\rangle$}}
\newcommand{\kthree}[3]{\mbox{$|{#1}\,,\,{#2}\,,{#3}\rangle$}}
\newfont{\blackb}{msbm10 scaled 1200}
\def\BR{\mbox{\blackb R}}
\def\BC{\mbox{\blackb C}}

\title{Quantum revivals, geometric phases and circle map recurrences}
\author{S. Seshadri,  S. Lakshmibala and V. Balakrishnan\\ 
{\em Department of Physics, Indian Institute of Technology - Madras,}\\ {\em
Chennai
600 036, India}} 
\date{}
\maketitle 
\begin{abstract}
 Revivals of the coherent states of a deformed, adiabatically
and cyclically varying 
 oscillator Hamiltonian are examined. 
The revival time distribution is exactly that
of Poincar\'{e} recurrences for a rotation map: 
only three distinct revival times can
occur, with specified weights. A  
link is thus established between quantum revivals and
recurrences in a coarse-grained discrete-time dynamical system.
\end{abstract} 
\mbox{} \hskip .85cm {\bf Keywords}: coherent state, wavepacket revival,
geometric phase, circle map,\\
\mbox{} \hskip .85cm  Poincar\'{e} recurrence
\newpage

\hskip .5cm Revivals  of a non-stationary state of a Hamiltonian may
be regarded   as the  quantum  analogs  of  recurrences  in a classical
dynamical system. In general, when  a quantum mechanical system is in
a state \kone{\psi}, the autocorrelation  function  $|\langle
\psi(0)\kone{\psi(t)}|^2$ decays from  its initial value  of unity with
increasing  time $t$. Under very  special circumstances,  however, it may
return to its initial value at certain  instants of time, signalling a
full revival. Typically, however, any return would be to a  value in a  small
neighborhood of unity \cite{ENS}, which we shall term a ``near
revival''. This would be the counterpart of
a Poincar\'{e} recurrence \cite{MK} in a
classical system, although, of course, there is no direct connection
between the revivals of quantum states and recurrences of the classical
phase trajectories of the same physical system.
Taking into account finite experimental resolution, it
is evident that near revivals rather than exact revivals are the readily
identifiable phenomena of interest.
Revivals and fractional revivals of wavepackets are of
considerable current interest in many contexts: atom optics \cite{RWK},
propagation of coherent light through a Kerr medium \cite{TAC}, Rydberg
states \cite{PS}, interacting atom-radiation
(Jaynes-Cummings) systems \cite{ENS, MVS, CBP} with single
and multi-photon coherent and
squeezed coherent states, quantum dynamics of boson systems and spin
systems \cite{Berman}, and so on. 

In this paper, we show that one can indeed establish a close
relationship between the near revivals of a wavepacket and recurrences in a
discrete-time dynamical system, by essentially mapping the former problem
onto the latter. This leads to a convenient and systematic way of
analyzing the statistics of revivals.  
In order to discretize time in an unambiguous and natural way, the
parameters  in the Hamiltonian  are varied adiabatically and  cyclically,
with a time period $T$. Consequently, 
it is only at instants separated by an interval $T$ that the Hamiltonian 
returns to its original self. It is therefore meaningful to look
for revivals of a state only at the instants $T, 2T, \cdots$.
In turn, this permits us to analyze the problem in terms of recurrences in
a discrete-time map. 

As a by-product of the cyclic variation of the parameters in the Hamiltonian,
the state vector may pick up an extra non-integrable (``geometric'')
phase \cite{WS}. One may expect the revival times to be affected as a
consequence -
{\em a priori}, to get shifted by the time required to cover the angular
excess (or decrement) representing the anholonomy in the semiclassical
limit, namely, the corresponding Hannay angle \cite{MVB, JHH}. This is indeed
the case \cite{CJ}, 
 and the specific system we work with yields an explicit verification of
this result as well. For definiteness, we consider a unitarily deformed
oscillator Hamiltonian whose eigenstates are squeezed generalized
coherent states, while the spectrum remains linear (equi-spaced). Coherent
states and
their generalizations \cite{AP} are convenient in this regard, as they
approximate
classical states of radiation. They also provide a setting in which  
anholonomies (non-integrable phases)
can be measured experimentally \cite{CW, PH} and analyzed
theoretically \cite{CSS, SLB},
in a physical range extending from the extreme quantum regime to the
semiclassical limit. Although revivals, {\em per se}, are of greatest
physical interest in the case of spectra that are nonlinear (in the
quantum numbers), here we  restrict our attention to a linear
spectrum. (In this case {\em exact} revivals correspond, as we shall see,
to a
trivial periodicity.) 
Complications such as fractional revivals \cite{AVP} do
not arise here, and this helps  us  establish in a clear
manner a very interesting link between near revivals,
anholonomies and recurrences -- manifestations, respectively, of quantum
interference, non-trivial topology in parameter space, and ergodicity in a
classical dynamical system.

We recall that the coherent state \kone{\zeta} ($\zeta  \in  \BC$) of
the standard oscillator Hamiltonian $H = \hbar \omega (\ad{a}a + 1/2)$
is  given by \kone{\zeta} = $D(\zeta) \, \kone{0}$ where \kone{0} is
the ground state of $H$, and $D(\zeta)$ is the displacement operator  
$\exp \,(\zeta \ad{a}-\zeta^* a)$. For each value of $\zeta$,
\kone{\zeta} is represented in the position basis by a Gaussian
wavepacket which does not spread with time (under evolution governed
by $H$). The centre of the wavepacket oscillates according to the
classical laws of motion. Using the expansion of \kone{\zeta(0)}
$\equiv \, \kone{\zeta}$ in terms of the eigenstates $\{ \kone{n}\}$
of $H$, its correlation function is easily found to be
\begin{eqnarray}
|\langle  \zeta(0)\kone{\zeta(t)}|^2 &=& \exp\,[\,2\,|\zeta|^2
\,(\cos\omega t-1)\,]\;. 
\end{eqnarray}
Thus, in this linear case, revivals   of the  initial wavepacket 
simply amount to periodicity with period $ 2\pi/ \omega$.
 To induce a geometric phase,
however, we need a Hamiltonian that has more than just a single parameter
(analogous to $\omega$ in $H$) that can be varied. Further, in order
to have a non-vanishing anholonomy (Hannay angle) in the
semiclassical limit, the geometric phase must have a dependence on
the quantum number $n$ labelling the states \cite{MVB}. Both these objectives
are
achieved 
if we begin with a deformation of $H$ produced by {\em squeezing}. For the
sake
of generality, we include a possible displacement as well, and define the
transformed Hamiltonian 
\begin{eqnarray}
{\widetilde H}(\alpha, \beta)  &=& S(\beta)\, D(\alpha)\, H \,
\ad{D}(\alpha) \, 
\ad{S}(\beta)\,\,
\end{eqnarray}
where $\alpha, \, \beta \in \BC$, and $S(\beta)$ is the squeezing
operator \cite{HPY} $S(\beta)\,=\, \exp\,((\beta {\ad{a}}^2 - \beta^*
a^2)$/2). As
$D(\alpha)$ and $S(\beta)$ are unitary, $H$ and $\widetilde {H}$ are
unitarily equivalent and isospectral. $\widetilde {H}$ has eigenstates
\kthree{n}{\alpha}{\beta} = $S(\beta)\, D(\alpha)\, \kone{n}$ ( $n = 0, 1,
\cdots$), connected by ladder operators $\tilde a$ and \ad{\tilde a}, where 
${\tilde a} =  S\,D\,a\, \ad{D} \, \ad{S}$ and [$\tilde a , \ad{\tilde a}]
=1$.

Under a cyclic, adiabatic variation of the displacement and squeezing
parameters $\alpha = \alpha_1 + i \, \alpha_2$ and $\beta = \beta_1 +i
\, \beta_2$, the state \kthree{n}{\alpha}{\beta} acquires
a Berry phase $\gamma_n$ that is a sum of contributions from the
displacement and squeezing parameters, respectively. 
The
contribution from the variation of the displacement parameter $\alpha$
can be found from the group multiplication law for the
elements $\{D(\alpha)\}$ of the Heisenberg-Weyl group, namely,
\begin{eqnarray}
D(\alpha)\, D(\alpha')&=& D(\alpha + \alpha') \, \exp \,\left [ \mbox{i}
\chi (\alpha, \alpha') \right ]
\end{eqnarray}
where $\chi(\alpha,\alpha')$ is twice the area of the triangle
with vertices at $0, \, \alpha'$ and $\alpha+\alpha'$ in the complex
plane. 
This contribution is independent of
$n$ essentially because $[a, \ad{a}]$ is just the identity operator. On the
other hand, the contribution from the variation of the squeezing
parameter $\beta$ is more involved. The squeezing operator $S(\beta)$
represents an element of the
group $SU(1,1)$, with generators $K_+ = \ad{a}^2 /2,\; K_{-} = a^2 /2$ and
$K_0 = [K_-,K_+]/2 = (a \ad{a}+\ad{a} a)/4$. The multiplication rule
is of the form \cite{AP}
\begin{eqnarray}
S(\beta)\, S(\beta') &=& S(\beta^{''}) \,\, \exp \, [\mbox{i}\, \phi
(\beta,\beta') \, K_0]
\end{eqnarray}
where $\beta^{''}$ is a certain function of $\beta$ and $\beta'$.  
The phase $\phi (\beta,
\beta')$ is related to the area of a certain geodesic triangle on the upper
sheet of the hyperboloid $\vec{x} \cdot \vec{x} \equiv x_0^2 - x_1^2
-x_2^2 =1$, where (setting $\mbox{arg}\, \beta \,=\, \varphi$) $\vec{x} =
(\cosh 2
|\beta|,\, - \sinh 2|\beta|\, \cos
\varphi,\, \sinh 2|\beta|\, \sin \varphi)$, on which the parameters of
$SU(1,1)$ live. The geometric phase 
is essentially
the solid angle subtended at the origin $\vec{x} = 0$ 
by the {\em invariant} area enclosed on the hyperboloid \cite{MS}. Since 
$\vec{x} \cdot \vec{x} =1$, this is just the surface integral 
over  $(dx_1 \,
dx_2)/x_0$ on the hyperboloid. 
In terms of $\beta$ this  integral is 
\begin{eqnarray}
\cal{B} &=& \int  \, d^2 \beta \,\, \frac{\sinh 2|\beta|}{|\beta|}\,\,,
\end{eqnarray}
the integration running over the area enclosed by the loop traversed in
the $\beta$-plane in a  cyclic variation of the squeezing parameters. 
Further, owing to the presence of the diagonal generator  
$K_0$ in Eq. (4), as opposed to the unit operator in the case of
$D(\alpha)$, there is an additional factor  $(n + \frac{1}{2})$ in the
geometric phase. Collecting these results,  
we arrive at the expression 
\begin{eqnarray}
\gamma_n &=& -2{\cal A} -  (n+\frac{1}{2}) {\cal B} \;,
\end{eqnarray}
where ${\cal A}$ is the area of the loop traversed in the $\alpha$-plane
in one cyclic variation of $\alpha_1$ and $\alpha_2$.
The linear dependence of $\gamma_n$ on $n$ \cite{SLB} implies that
${\cal A}$ and 
${\cal B}$ are determined by $\gamma_0$ and $\gamma_1$ (and vice
versa): $ {\cal A} \,=\, (\gamma_1 - 3\gamma_0)/4$,  ${\cal B} =
\gamma_0 - \gamma_1$.

We now construct the coherent state
\kone{z} {\em in the tilde-basis}, i.e., $\tilde a \kone{z} = z
\kone{z}\;\; (z \in \BC)$, so that 
\begin{eqnarray}
\kone{z} &=& \exp\, \left (-\frac{|z|^2}{2} \right ) \,
\sum_{n=0}^{\infty}
\,\,\frac{z^n}{\sqrt{n!}} \,\,  \kthree{n}{\alpha}{\beta} \;\;. 
\end{eqnarray}
In the present instance, an equivalent way of obtaining \kone{z} is  by
operating on the ground state \kthree{0}{\alpha}{\beta} by the
displacement operator $\exp (z \ad{\tilde a}- z^{*} \tilde{a})$.
Consider time evolution governed by the Hamiltonian ${\widetilde H}$,
while ${\widetilde H}$ itself changes adiabatically owing to a cyclic
variation of the parameters $\alpha$ and $\beta$ with a time period $T
>> \omega^{-1}$. An initial state \kone{z(0)} $\equiv$ \kone{z} (given
by Eq. (7)) evolves at time $T$ to  
\begin{eqnarray}
\kone{z(T)} &=& \exp\, \left (-\frac{|z|^2}{2} \right ) \,
\sum_{n=0}^{\infty}
\frac{z^n}{\sqrt{n!}}\,
\exp \, \left [-i \left (\gamma_n - (n+\frac{1}{2})\,\omega T\right )
\right ] \, \kthree{n}{\alpha}{\beta} \;\;.
\end{eqnarray}
Using the results quoted in Eqs. (5) and (6) for $\gamma_n\,$, the
correlation function now works out to
\begin{eqnarray}
|\langle z(0) \kone{z(T)}|^2 &=& \exp\left [\, 2|z|^2\, (\cos(\omega T +
{\cal B}) -1)\,\right ]\;.
\end{eqnarray}
Therefore revivals occur at times $T,$ $2T, \cdots$ {\em provided}
$\omega T + {\cal B} = 2 \pi p$, where $ p=0,1, \cdots$ -- that is,
provided the cyclic variation of the parameters in the Hamiltonian is
carried out in a time period equal to one of the values $(2\pi - {\cal
B}) / \omega ,\; (4\pi - {\cal B}) / \omega,\, \cdots$. This is to
be compared with the original revival times $2 \pi p / \omega$
(cf. Eq. (1)). As  $- \partial \gamma_n / \partial
n\;=\; {\cal B} \; (=\, \gamma_0 - \gamma_1$ in the present instance)
is just  the Hannay angle \cite{MVB}, we have here an explicit verification
of the
staggering of the  revival times  by precisely the time required to cover  
this angular excess (or decrement, depending on its sign) \cite{CJ}.

Next, we note that there is a
convenient and natural way to describe the evolution of
the initial state \kone{z(0)} to the state at times $T, 2T, \cdots\,$.
Equation (8) can be re-written as 
\begin{eqnarray}
\kone{z(T)} &=& \exp \, \left (i\,(\gamma_0 - \frac{1}{2} \omega T\,)
\right ) \,\,\kone{z\, e^{-i(\omega T + {\cal B})}}\;.
\end{eqnarray}
Moreover, as the cyclic variation of $\alpha$ and $\beta$ is
continued, the state \kthree{n}{\alpha}{\beta} picks up the same
additional phase $\gamma_n$ in {\em each} cycle -- i.e., the same
Hannay angle ${\cal B}$ is added in each interval $T$. Therefore,
if $\theta_k$ denotes the phase of the complex eigenvalue labelling
\kone{z(kT)} ($k = 0,1,2, \cdots$), the evolution is  equivalent
to a rotation map \cite{RLD} on a circle ($S^1$), namely,
\begin{eqnarray}
\theta_{k+1} &=& \theta_k - 2 \pi \Delta  \;\;\; \mbox{(mod 2$\pi$)}
\end{eqnarray}
where $ 2 \pi \Delta = \omega T + {\cal B}$. If $\Delta$ is an integer,
every value of $\theta$ is periodic, with period 1. This is the case
already discussed following Eq. (9). A more general possibility is $\Delta
= p/q$, a rational number. Again, every orbit of the map is periodic, but
with a period $q$. Correspondingly, every initial state \kone{z(0)} has
revivals at times $qT,\; 2qT,\; \cdots\;$.  But rational values of
$\Delta$ constitute a set of measure zero. The generic case corresponds to
an irrational value of $\Delta$, and this is also the most interesting 
case. (It is also the most pertinent one from a practical point of view, 
as it relates directly to near revivals, as we shall see.) As is
well known, the map no longer has any periodic orbits, but the iterates of
{\em any} $\theta_0$ cover $S^1$ densely as $k \rightarrow \infty$.
(Regarded as the Poincar\'{e} section for  motion on a 2--torus, this is
the quasi-periodic case). The dynamics is ergodic but not mixing, with a
uniform invariant density $\rho(\theta) = 1/(2\pi)$. Although we no longer
have {\em exact} revivals in principle, the system comes arbitrarily close
to these in practice ({\em near} revivals) -- precisely the analog of
Poincar\'{e} recurrences. As a consequence, the following  complete
analysis of the statistics of near revivals becomes possible. 

Consider a prescribed small angular interval $I_{\epsilon}$ of size $2\pi
\epsilon$,
located symetrically about the initial phase $\theta_0$ on $S^1$. Let
$\Delta \equiv [\Delta] + \, \delta$, where $[\Delta]$ stands for the
integer defined by 
$\Delta -1 < [\Delta] < \Delta$ (for {\em either} sign of $\Delta$), so
that
$\delta$ is an irrational number satisfying $0< \delta <1$. It is then
easy to see that  $\theta_k
\in I_{\epsilon} \Rightarrow \{k\delta\} < \epsilon$, where $\{ x \}$ denotes
the fractional part of $x$. The corresponding correlation function then
merely differs from unity by a term of order $\epsilon^2$, because 
\begin{eqnarray}
|\langle z(0)\kone{z(kT)}|^2 &=&\exp\, \left [2|z|^2 \,(\cos \,2\pi k
\Delta -1)\right ] \nonumber \\
\,&>& \; 
\exp \, \left (\, 4|z|^2 \,\pi^2 \{ k \delta\}^2\right ) \;>\; 1-4|z|^2
\,\pi^2 \, \epsilon^2\,.
\end{eqnarray}
To order $\epsilon$, therefore, we may regard  a return of $\theta_k$ to
$I_{\epsilon}$ as a (near) revival. The statistics of the occurrence times
of such revivals  is then
identical to that of the recurrences to an angular interval of size $2\pi
\epsilon$ in the rotation map (11). The solution to the latter problem is
given by certain gap theorems for interval exchange transformations
\cite{GR, NBS}.
Applying these  to the case at hand, we obtain the following results 
in the long-time limit, after the transients due to specific initial
conditions have died out and the invariant measure is attained.     

Ergodicity implies that the {\em mean} recurrence time (here, the mean
time
between successive near revivals) is, in units
of
$T$, the reciprocal of the invariant measure of $I_{\epsilon}$. As
this measure is uniform on the circle, the mean recurrence time is just 
$T/ \epsilon$, as one may expect. The {\em distribution} of recurrence
times  is, however, quite remarkable \cite{N4}. In general (i.e., for
arbitrary $\epsilon$ and $\delta$), it is
concentrated at no more than {\em three} points $T_1 = k_1 T, \; T_2 = k_2
T$ 
and $T_3 = T_1 + T_2$, where $k_1$ and $k_2$ are the least positive
integers such that \begin{equation} \{k_1 \delta \} \;<\; \epsilon \;\;
\mbox{and} \;\; 1-\{k_2 \delta\} \; < \; \epsilon \end{equation}
respectively.  (Recall that $\epsilon \ll 1$; as long as $\epsilon <
1/2,\; k_1 \neq k_2$.)  It is easy to show that $\epsilon \leq \{k_1
\delta \} + 1 - \{ k_2 \delta \}$: when the equality sign applies, only
{\em two} recurrence times ($T_1$ and $T_2$) occur. If $F(kT)$ denotes the
normalized invariant (i.e., post-transient) probability that the revival
time is $kT$ ($k$ = positive integer), then 
\begin{eqnarray}
F(kT) &=&   (1/ \epsilon)\, \left [ \,\frac{}{} \left (\epsilon - \{k_1
\delta \}
\right ) \delta_{k,k_1}
 + \left (\epsilon - 1 + \{k_2 \delta \}\right ) \delta_{k,k_2}  
\right. \nonumber \\  
  & &  \left. \mbox{    }\frac{}{} +  \left (\,\{k_1 \delta \}+ 1 - \{k_2
\delta
\} -
\epsilon\, \right ) \, \delta_{k,k_1+k_2}\, \right ] \;. 
\end{eqnarray}
This is consistent with the requirement $\langle k \rangle = 1/ \epsilon$
(which follows from ergodicity) provided $k_2\, \{k_1 \delta \} + k_1 (1 -
\{k_2 \delta \}) \,=1$, a relationship that can be established
independently. With slight modifications, these results for near revivals
continue to hold good in the periodic case
($\delta $ = rational number) as well. {\em They are therefore generic,
and do not require any fine-tuning of the parameters in the problem}
($\omega, \; T $ and ${\cal B}$). 

Revivals are a manifestation of the interference arising from the
different phases acquired under time evolution by the stationary basis 
states in the expansion of a non-stationary state. Hence it is the 
quantum-number-dependent part of the phase (both dynamical and geometric)
that is relevant in this regard. That is why, in the present instance, it
is only the coefficient ${\cal B}$ in the expression for $\gamma_n$ (see
Eq.
(6))  that plays a role, and not  the quantity ${\cal A}$. (We
have already seen why varying $\alpha$ does not lead to an
$n$-dependent geometric phase, while varying $\beta$ does so.) Therefore,
as far as revivals are
concerned,  we may hold $\alpha$ fixed at the value zero throughout
(i.e., displacement  may be dispensed with altogether),
without altering any of the
conclusions. A cyclic variation of the squeezing parameters $\beta_1$ and
$\beta_2$  suffices
to produce the relevant non-trivial anholonomy. 
For any given values of $\omega T$ ($\gg 1$) and ${\cal B}$, the
near revivals
of the correlation function concerned have been shown to be essentially
equivalent  to the Poincar\'{e} recurrences in a rotation map on $S^1$. 
The steady-state distribution of recurrence times is explicitly
determined. It is restricted to just three possible values. Their
locations and relative frequencies of occurrence are easily found for any
given $\delta$ and prescribed $\epsilon$.

As already stated, we have considered near revivals in the case of a 
linear spectrum. In the general nonlinear case
(including, in
the quantum optical context, multiphoton coherent states \cite{SCSA}),
additional
interesting features appear, that have to be taken into account in
establishing a mapping between near revivals (allowing for the effects of 
possible non-integrable phases) and 
Poincar\'{e}  recurrences. These include fractional revivals \cite{AS} and
nonlinear dependence of the geometric phase on the quantum numbers. 
Details of this investigation  will be reported elsewhere.

SS acknowledges financial support from the Council of Scientific and 
Industrial Research, India in the form of a Senior Research Fellowship. SL
thanks the Department of Science and Technology, India, for partial
support under the grant SP/S2/E-03/96. We are grateful to S. Govindarajan
for helpful discussions and to the referees for their suggestions.

\end{document}